# Comparison of the material quality of $Al_xIn_{1-x}N$ ($x \sim 0$-$0.50$) films deposited on Si (100) and (111) by reactive RF sputtering

M. Sun[1*], R. Blasco[1,2], M. de la Mata[3], S.I. Molina[3], A. Ajay[4], E. Monroy[4], S. Valdueza-Felip[1] and F. B. Naranjo[1]

[1]Photonics Engineering Group. Electronics Department, University of Alcala, Madrid-Barcelona road km 33.6, 28871 Alcalá de Henares, Spain.
[2]Science, Computation and Technology Department, European University of Madrid, Tajo street, s/n, 28670, Villaviciosa de Odón, Madrid, Spain.
[3]Dpto. Ciencia de los Materiales, I. M. y Q. I., IMEYMAT, Universidad de Cádiz, Campus Río San Pedro, s/n, 11510, Puerto Real, Cádiz, Spain.
[4]CEA-Grenoble, INAC/PHELIQS, 17 av. des Martyrs, 38054, Grenoble, France.

*michael.sun@uah.es

$Al_xIn_{1-x}N$ ternary semiconductors have attracted interest for application in solar cell devices. Here, we compare the material quality of $Al_xIn_{1-x}N$ layers deposited on Si with different crystallographic orientations, (100) and (111), via radio-frequency (RF) sputtering. To modulate their Al content, the Al RF power was varied from 0 to 225 W, whereas the In RF power and deposition temperature were fixed at 30 W and 300ºC, respectively. X-ray diffraction measurements reveal a *c*-axis-oriented wurtzite structure with no phase separation regardless of the Al content ($x = 0$-$0.50$), which increases with the Al power supply. The surface morphology of the $Al_xIn_{1-x}N$ layers improves with increasing Al content and it is similar for samples grown on both Si substrates (the root-mean-square roughness decreases from $\approx 12$ nm to 2.5 nm). Furthermore, from TEM images we notice a similar grain-like columnar morphology and defect density on samples deposited on both Si substrates under the same conditions. Simultaneously grown $Al_xIn_{1-x}N$-on-sapphire samples point to a residual *n*-type carrier concentration in the $10^{20}$-$10^{21}$ cm$^{-3}$ range. The optical band gap energy of these layers evolves from 1.75 eV to 2.56 eV with increasing Al content, consistent with the blue shift of their low-temperature photoluminescence. In general, the material quality of the $Al_xIn_{1-x}N$ films on Si is similar for both crystallographic orientations. Nonetheless, samples deposited on sapphire show an improved structural and morphological characteristic likely due to the lower difference in lattice constants between the nitride and the sapphire substrate.

# 1. INTRODUCTION

$Al_xIn_{1-x}N$ ternary semiconductor alloys have attracted huge interest for their application in solar cell devices, particularly after the revision of the indium nitride band gap in 2001 [1]. The direct band gap (i.e. high absorption coefficient) of $Al_xIn_{1-x}N$, tunable from the near infrared (0.7 eV for InN [1]) to the ultraviolet (6.2 eV for AlN [2]), makes it an excellent candidate for developing photovoltaic devices in combination with silicon. In addition, this material shows a high resistance to thermal/mechanical stress and irradiation with high-energy particles [3], which makes it suitable for space applications.

The synthesis of high-quality single-phase $Al_xIn_{1-x}N$ layers is challenging due to the large difference in properties like bonding energy, lattice constants or growth temperature between the binary constituents, InN and AlN. The growth of $Al_xIn_{1-x}N$ layers has been reported by various techniques, including metal-organic chemical vapor deposition (MOCVD) [4-7], molecular beam-epitaxy (MBE) [8-12], and reactive sputtering. Within this last technique we can distinguish two approaches, one that uses a mixture of argon and nitrogen for the deposition [13-22] and another that uses only nitrogen [23-29]. Reactive sputtering allows the deposition on large substrates and employs lower temperatures than MOCVD or MBE. However, the low-temperature deposition comes at the price of higher defect density. The presence of impurities like hydrogen [30] and defects like nitrogen vacancies [31] can induce unintentional doping with a residual carrier concentration as high as $10^{21}$ cm$^{-3}$, which causes a blue shift of the optical band gap due to the Burstein-Moss effect [32].

$Al_xIn_{1-x}N$ can be synthesized on different substrates, such as Si(111) [20,28,29,33,34,35], Si(100) [17,21,25,36], sapphire [15,20,24,28,33,35,36], glass [15,20,28,35] or GaAs [20]. However, the properties of the $Al_xIn_{1-x}N$ films depend strongly on the nature of the substrate. It is particularly interesting to study the deposition on silicon due to its potential for hybrid solar cells. Wurtzite III-nitrides are usually grown on silicon (111) due to the hexagonal symmetry of this crystallographic plane. However, today silicon-based nanotechnology uses silicon (100) due to the lower amount of dangling bonds, which generate undesired recombination centers [32].

There are several studies about the growth of $Al_xIn_{1-x}N$ films either on Si(111) or on Si(100), and compared to $Al_xIn_{1-x}N$ on sapphire substrates. Bashir *et al.* [33] deposited InN on Si(111) by RF sputtering, obtaining large crystallite size, low micro-strain and low dislocation density. Afzal *et al.* [20] grew $Al_xIn_{1-x}N$ films on Si(111) at 300ºC using a magnetron co-sputtering system obtaining polycrystalline films with preferred orientation along the (101) direction, with higher crystallite size and lower surface roughness compared to other substrates like GaAs and glass. However, the comparison of $Al_xIn_{1-x}N$ layers simultaneously grown on both Si(111) and Si(100) substrates by reactive RF sputtering is missing yet.

Here, we compare ternary $Al_xIn_{1-x}N$ layers with varying Al content (0-50% Al) deposited simultaneously on silicon (100) and (111) via reactive RF sputtering, in terms of their structural, morphological, electrical and optical properties.

## 2. Experimental section

$Al_xIn_{1-x}N$ layers were deposited in a reactive RF magnetron sputtering system (AJA International, ATC ORION-3-HV) simultaneously on three substrates: *p*-doped 375-µm-thick Si(100), *p*-doped 500-µm-thick Si(111) (both with a resistivity of 1-10 Ωcm) and on 500-µm-thick (0001)-oriented sapphire. This system is equipped with 2-inch confocal magnetron cathodes of pure In (4N5) and pure Al (5N). The base pressure of the system was in the order of $10^{-7}$ mbar. The substrate-target distance was fixed at 10.5 cm and the temperature during the deposition was monitored with a thermocouple placed in direct contact with the substrate holder. The substrates were chemically cleaned in organic solvents before being loaded in the chamber, where they were outgassed for 30 min at 550°C, and then cooled down to the growth temperature. Previous to the deposition, the surface of the targets and the substrates were cleaned using a soft plasma etching with Ar (2 sccm and 20 W). $Al_xIn_{1-x}N$ layers were deposited in a pure $N_2$ atmosphere with a nitrogen flow of 14 sccm and a pressure of 0.47 Pa. The RF power applied to the Al target, $P_{Al}$, was set to 0, 100, 125, 150, 175 and 225 W (samples M1-M5, respectively) while the RF power applied to the In target and the temperature were fixed to 30 W and 300°C, respectively. A sputtering time of 3 hours was used for the InN sample, 5 hours for the sample with $P_{Al}$ = 100 W and 4 hours for the rest. The thickness and deposition rate of the samples are summarized in Table 1.

The alloy mole fraction, crystalline orientation and mosaicity of the films were evaluated by high-resolution X-ray diffraction (HRXRD) measurements using a PANalytical X'Pert Pro MRD system. In addition, the thicknesses of the layers were obtained from field-emission scanning electron microscopy (FESEM) images. Atomic force microscopy (AFM) was employed to study the surface morphology and estimate the root-mean-square (rms) surface roughness using a Bruker multimode Nanoscope IIIA microscope in tapping mode. Additionally, transmission electron microscopy (TEM) provided a deeper understanding of the structural properties of the interface between the deposited material and the substrate. The electrical properties of the films were analyzed using room temperature Hall-effect measurements in a conventional Van der Paw geometry.

Optical transmission measurements were carried out on the samples grown on sapphire. The measuring system consists of a white broadband lamp whose emission is collimated to obtain a homogenous beam. The sample was placed in the optical path of this beam, and a 20× magnification microscope objective focused the transmitted light into a 600-um diameter optical fiber connected directly to an optical spectrum analyzer (OSA) with two detectors (Si and InGaAs) that cover the visible and the near-infrared spectral range (350 to 1700 nm). All the spectra were corrected by the emission spectrum of the lamp.

Finally, photoluminescence measurements were carried out by exciting the samples with ~30 mW of a continuous-wave laser diode emitting at λ = 405 nm focused onto a 1 mm diameter spot. The emission was collected with a 193-mm focal-length Andor spectrograph equipped with a UV-extended silicon-based charge-coupled-device camera operating at −60°C between 200 nm and 1100 nm.

## 3. Results and discussion

### 3.1. Structural characterization

To study the structural quality of the layers, HRXRD 2θ/ω scans were carried out on the $Al_xIn_{1-x}N$ layers grown on Si(100), Si(111) and sapphire, with the results shown in Figures 1(a), 1(b) and 1(c), respectively. All layers present wurtzite crystalline structure highly oriented along the *c*-axis, and no other crystallographic phases are detected. The increase in $P_{Al}$ shifts the (0002) and (0004) reflection peaks assigned to $Al_xIn_{1-x}N$ towards higher diffraction angles, which confirms the reduction of the *c* lattice parameter. The Al mole fraction of the alloy was estimated applying the Vegard's Law [38] to the AlN-InN system, using the *c* lattice parameter obtained from HRXRD and assuming fully relaxed layers. The calculated Al mole fraction, *x*, scales linearly with $P_{Al}$ in the range of $x = 0$ to $x = 0.49, 0.48, 0.43$ for Si(100), Si(111) and sapphire substrates, respectively, as summarized in Table 1.

The FWHM of the ω-scan (rocking curve) of the (0002) $Al_xIn_{1-x}N$ diffraction peak provides information about the mosaicity of the material. In this study, $Al_xIn_{1-x}N$ layers grown on both silicon substrates show similar values, in the 3-6º range, without clear trend (see Table 1). This indicates that the mosaicity is independent of the crystal orientation of the silicon substrate. However, $Al_xIn_{1-x}N$ layers deposited on sapphire show significantly lower FWHM rocking curve (<1.9º, see Table 1), which is explained by the lower mismatch of lattice parameter and thermal expansion coefficient between III-nitrides and sapphire in comparison to silicon [39].

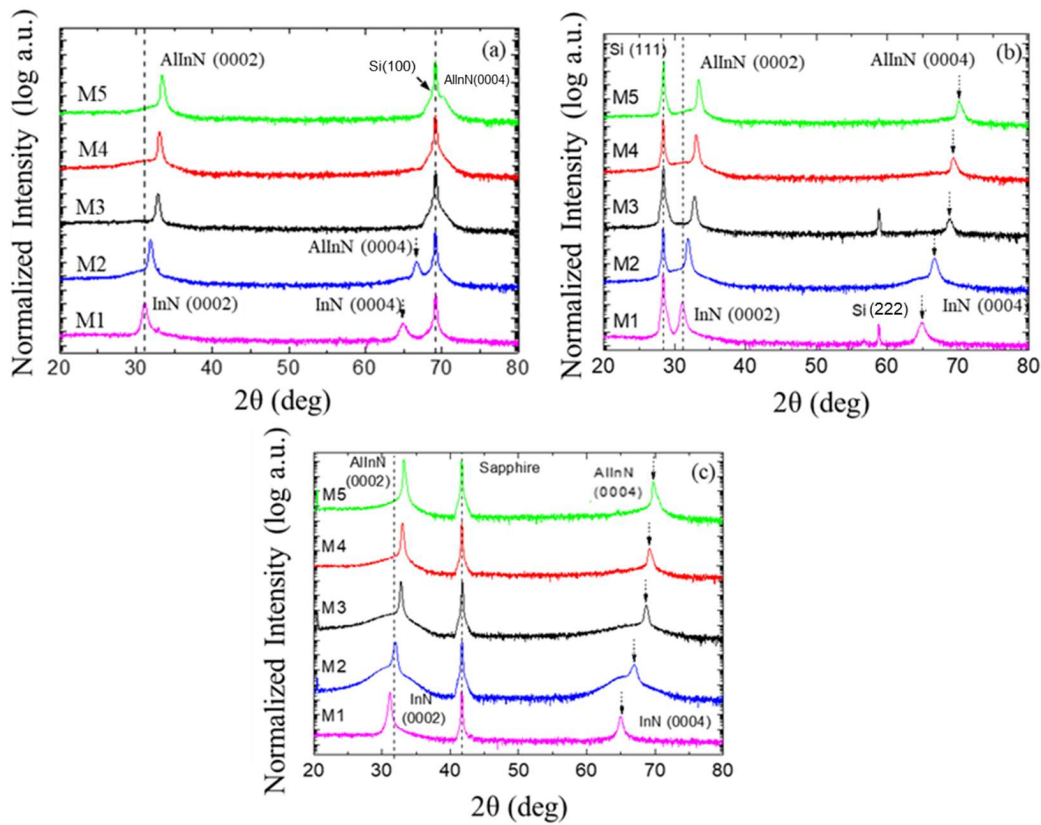

*Figure 1*: 2θ/ω scans of the $Al_xIn_{1-x}N$ layers deposited on (a) Si(100), (b) Si(111) and (c) sapphire for different $P_{Al}$. The only reflections assigned to $Al_xIn_{1-x}N$ are (0002) and (0004). The rest of the reflections are assigned to the substrates.

## 3.2. Morphological properties

In order to investigate the morphology of the layers, they were studied by FESEM and AFM. Figure 2(a), (b) and (c) shows FESEM images of samples grown on Si(111), Si(100) and sapphire, respectively. The morphology of the layers grown on silicon evolves from nanocolumnar for pure InN (M1) towards grain-like compact layers for increasing Al contents (M3 and M5). Such trend has been already observed in similar $Al_xIn_{1-x}N$ samples deposited on Si(111) by RF sputtering (40W In, 300ºC) with similar Al compositions [29]. The observed phenomena can be attributed to changes in the surface diffusion of adatoms due to the increased kinetic energy of the incoming Al species.

| Sample | Substrate | $P_{Al}$ (W) | $c$ (Å) | Al mole fraction $x$ | FWHM rocking curve (°) | Thickness (nm) | Deposition rate (nm/h) | Rms surface roughness (nm) |
|---|---|---|---|---|---|---|---|---|
| M1 | Si(100) | 0 | 5.73 | 0 | 4.6 | 376 | 130 | 11.6 |
| M2 | | 100 | 5.61 | 0.12 | 2.4 | 746 | 150 | 9.3 |
| M3 | | 150 | 5.45 | 0.35 | 6.2 | 563 | 140 | 3.5 |
| M4 | | 175 | 5.42 | 0.40 | 3.2 | 550 | 140 | 3.6 |
| M5 | | 225 | 5.36 | 0.48 | 2.8 | 714 | 180 | 2.4 |
| M1 | Si(111) | 0 | 5.73 | 0 | 4.7 | 368 | 120 | 12.8 |
| M2 | | 100 | 5.59 | 0.16 | 2.9 | 703 | 140 | 7.9 |
| M3 | | 150 | 5.44 | 0.36 | 6.1 | 618 | 160 | 3.6 |
| M4 | | 175 | 5.40 | 0.42 | 3.1 | 535 | 130 | 3.7 |
| M5 | | 225 | 5.35 | 0.49 | 2.8 | 757 | 190 | 2.6 |
| M1 | Sapphire | 0 | 5.73 | 0 | 1.9 | 247 | 80 | 6.0 |
| M2 | | 100 | 5.60 | 0.14 | 1.9 | - | - | 3.0 |
| M3 | | 150 | 5.47 | 0.32 | 1.4 | 322 | 80 | 0.6 |
| M4 | | 175 | 5.43 | 0.38 | 1.4 | 357 | 90 | 0.5 |
| M5 | | 225 | 5.39 | 0.43 | 1.1 | 564 | 140 | 0.4 |

*Table 1*: Summary of the deposition parameters, and the structural and morphological analysis of $Al_xIn_{1-x}N$ on Si(100), Si(111) and sapphire substrates: *c*-axis parameter and Al mole fraction *x* extracted from HRXRD, layer thickness estimated from FESEM, and rms surface roughness measured by AFM.

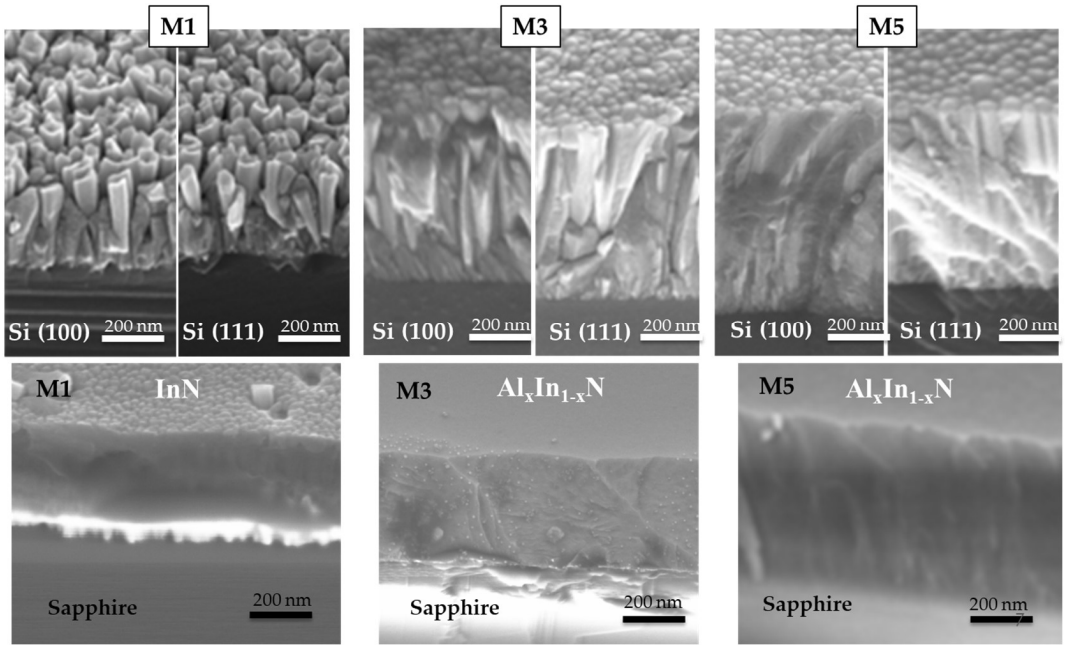

*Figure 2*: FESEM images of Al$_x$In$_{1-x}$N samples (a) M1, (b) M3, and (c) M5 on Si(100) (top left), Si(111) (top right) and sapphire (down).

The observed morphological transition is accompanied by a modification of the sample surface. The rms roughness was measured by AFM, scanning a 2×2 µm$^2$ area (Figure 3). The results shows a surface roughness evolution from 11.6 (Al content *x* = 0) to 2.4 (*x* ≈ 0.36) nm for Si(100) and from 12.8 (Al content *x* = 0) to 2.6 nm (*x* ≈ 0.36) for Si(111), in agreement with previous results [29]. Thus, the roughness remains almost constant for samples with an Al content in the range within *x* ≈ 0.36 - 0.42 (see Table 2), and it finally drops further up to the minimum for an Al content around 0.5.

The layers deposited on sapphire substrate show a completely different morphology, which is compact across the entire compositional range analyzed. This morphology is accompanied with smooth surfaces and very small values of rms surface roughness, below 1 nm for Al contents above 0.32.

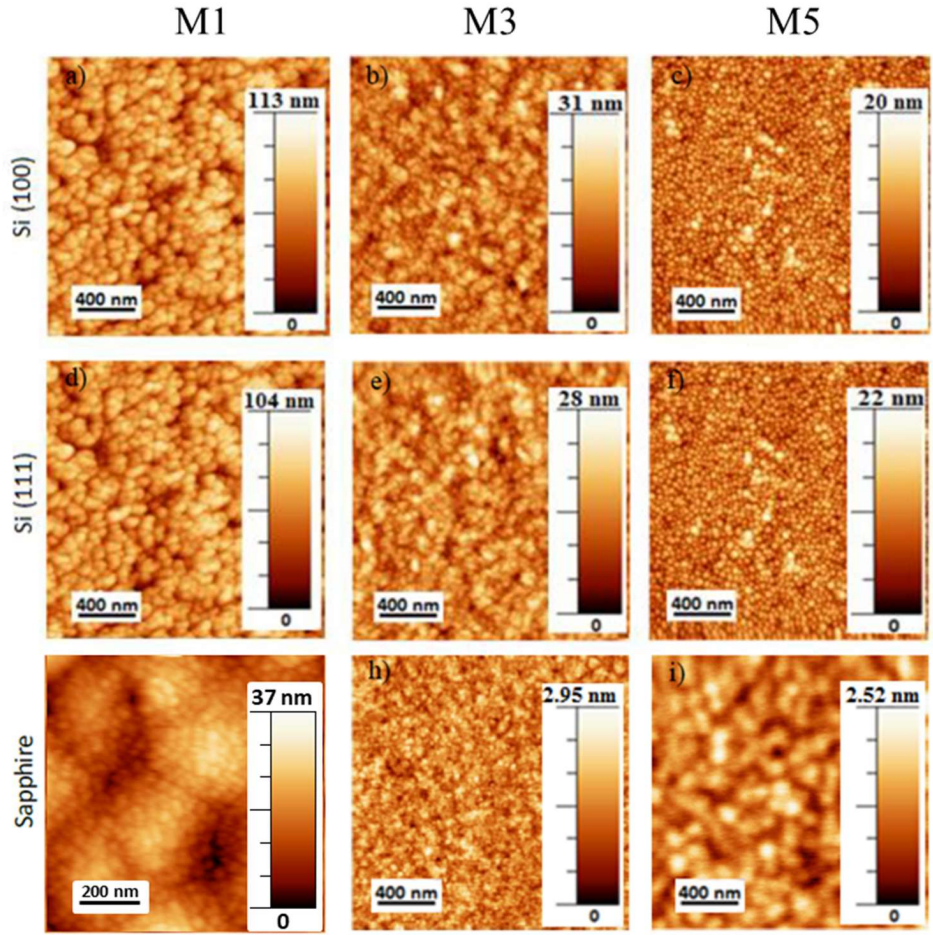

*Figure 3*: AFM images scanning an area of 2×2 μm² of InN and $Al_xIn_{1-x}N$ samples with $P_{Al}$ = 0, 150 and 225 W grown on Si(100) (a, b and c), Si(111) (d, e and f) and sapphire (g, h and i). From left to right, samples M1, M3 and M5.

In addition, samples of $Al_xIn_{1-x}N$ deposited on Si(100) and Si(111) under the same growth conditions as sample M3 ($P_{In}$ = 30 W, $P_{Al}$ = 150 W) but at 550ºC were investigated by transmission electron microscopy (TEM) to address their structural quality and presence of defects. As it was demonstrated by Núñez-Cascajero *et. al.* [29], there are no significant differences in the structural and morphological characteristics between $Al_xIn_{1-x}N$ layers deposited at 300ºC and 550ºC on Si(111) by RF sputtering with this Al content ($x \approx 0.36$). Figure 4 shows the similar grain-like columnar morphology of samples deposited on both substrates, Si(100) (a) and Si(111) (b). The AlInN wurtzite grains (pointed by black arrows in Figure 4) grow along the c-axis (aligned with the substrate) but showing slight in-plane rotation between neighbor grains. We can infer a similar defect density given the similar shape and dimensions of the grains (between 20 and 70 nm in projected diameter,) grown on both substrates, Si(111) and Si(100).

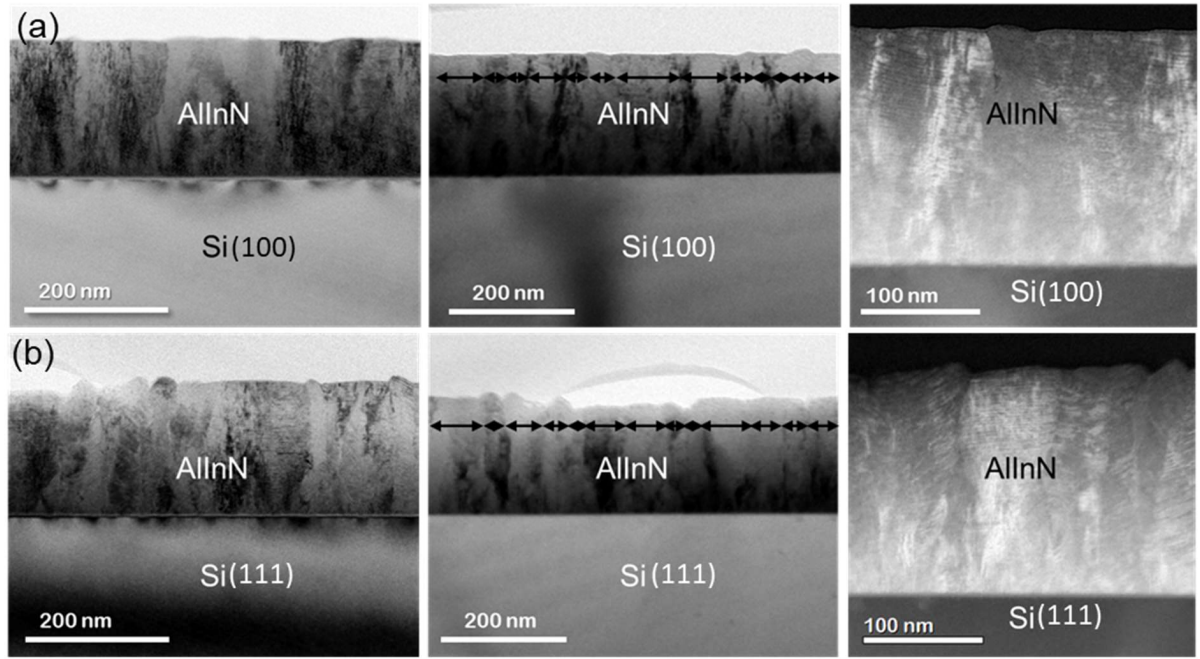

*Figure 4*: TEM images at different magnifications of $Al_{0.36}In_{0.64}N$ samples grown at 550ºC on (a) Si(100) and (b) Si(111) substrates. Right column images are displayed with inversed contrast to ease the visualization of distorted lattice regions (brighter).

### 3.3. Electrical characterization

The electrical properties of the $Al_xIn_{1-x}N$ layers could only be addressed for samples deposited on sapphire substrates, since the silicon conduction masked the layer signal whenever a silicon substrate is used.

Table 2 presents a summary of the Hall effect measurements performed in $Al_xIn_{1-x}N$ layers grown on sapphire. The layer resistivity increases from 0.38 mΩ·cm for InN to 8 mΩ·cm for $Al_{0.32}In_{0.68}N$, while the carrier concentration decreases from $1.73 \times 10^{21}$ cm$^{-3}$ for InN to $2.48 \times 10^{20}$ cm$^{-3}$ for $Al_{0.32}In_{0.68}N$, respectively. The values of resistivity and mobility obtained for the $Al_{0.32}In_{0.68}N$ sample are similar to those reported by Liu *et al.* [36] (1.2 mΩ·cm and 11.4 cm$^2$/V·s for a ~90 nm $Al_{0.28}In_{0.72}N$ layer deposited by RF sputtering at 600ºC). The high carrier concentration of the layers has been related to the unintentional doping from impurities such as hydrogen or oxygen during growth [30][49] and it was also observed by Nuñez-Cascajero *et al.* [24], where similar $Al_xIn_{1-x}N$ on sapphire with an homogeneous distribution of oxygen were obtained. Samples with an Al content above *x* = 0.32 show a resistivity above 10 mΩ·cm making the Hall effect measurement unreliable.

| Sample | Al mole fraction x | Resistivity (mΩ·cm) | Carrier concentration (cm$^{-3}$) | Mobility (cm$^2$/V·s) |
|---|---|---|---|---|
| M1 | 0 | 0.38 | $1.7 \times 10^{21}$ | 9.5 |
| M2 | 0.14 | 0.58 | $9.4 \times 10^{20}$ | 11.5 |
| M3 | 0.32 | 8.00 | $2.5 \times 10^{20}$ | 3.2 |

*Table 2*: Summary of the electrical characterization of $Al_xIn_{1-x}N$ layers deposited on sapphire obtained by Hall-Effect measurements at room temperature.

Figure 5 shows the evolution of the carrier concentration (filled squares) and resistivity (filled triangles) of our $Al_xIn_{1-x}N$ samples as a function of the Al mole fraction compared to values obtained by Nuñez-Cascajero *et al.* in $Al_xIn_{1-x}N$ on sapphire samples deposited under similar growth conditions on sapphire [24] (open squares and triangles).

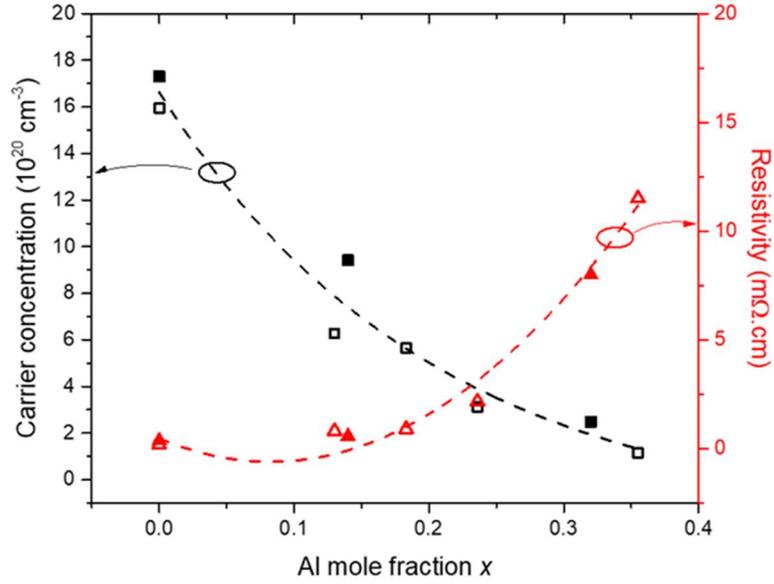

*Figure 5*: Evolution of the carrier concentration (filled squares) and resistivity (filled triangles) as a function of Al mole fraction compared to values obtained by Nuñez-Cascajero *et al.* [24] (open squares and triangles).

### 3.4. Optical properties

#### 3.4.1 Transmittance measurements

The optical properties of layers deposited on sapphire were measured by room-temperature optical transmittance. The inset of Figure 6 shows the transmittance spectra of all samples under study. The absorption of the layers ($\alpha$) can be derived from transmission spectra (*T*) following the relation $\alpha(E) \cdot L = -ln(T)$, being *L* the thickness of the layer, and without considering optical scattering and reflection losses. For the analysis, the obtained absorption coefficient has been fitted with the sigmoidal approximation [41] $\alpha(E) = \dfrac{\alpha_0}{1 + e^{\frac{E_{g,eff} - E}{\Delta E}}}$, where $\alpha_0$ is a coefficient that represents the absorption coefficient well above the band gap energy, $E_{g,eff}$ is the effective band gap energy, *E* is the photon energy, and $\Delta E$ is the absorption band edge broadening. The apparent optical band gap energy of the samples is then calculated from the linear fit of the squared absorption coefficient ($\alpha^2$) as a function of the photon energy (see Figure 6). As expected, the apparent optical band gap energy blue shifts with the Al mole fraction from $E_g^{Abs}$ ~ 1.73 eV (716 nm) for InN (M1) to $E_g^{Abs}$ ~ 2.56 eV (483 nm) for $Al_{0.43}In_{0.57}N$ (M5). See Table 3 for all optical data.

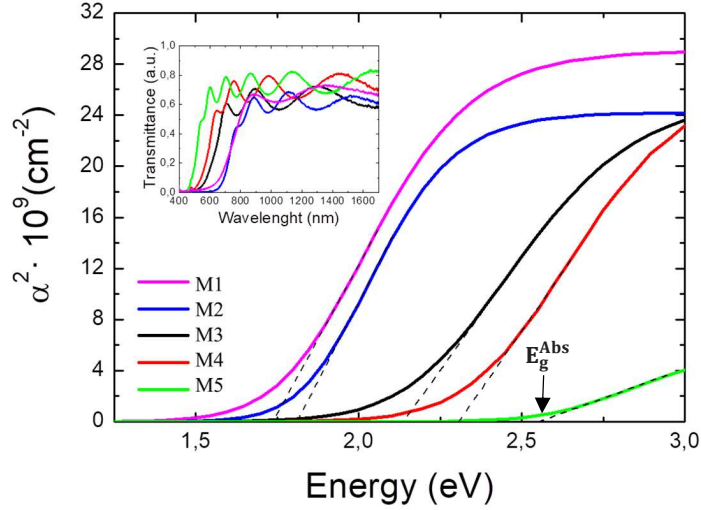

*Figure 6*: Squared absorption coefficient $\alpha^2$ as a function of the energy extracted from the sigmoidal approximation of the $Al_xIn_{1-x}N$ layers grown on sapphire. Dashed lines are the linear fits used to estimate the apparent optical band gap energy of the samples $E_g^{Abs}$. Inset: transmittance spectra vs wavelength of the same samples M1-M5.

| Sample | Al mole fraction $x$ | $\alpha_0 (\times 10^4 \text{ cm}^{-2})$ | $E_g^{Abs}$(eV) | $\Delta E$(meV) |
|---|---|---|---|---|
| M1 | 0 | 17.2 | 1.73 | 160 |
| M2 | 0.14 | 20.3 | 1.79 | 117 |
| M3 | 0.32 | 18.4 | 2.13 | 214 |
| M4 | 0.38 | 18.3 | 2.27 | 207 |
| M5 | 0.43 | 10.0 | 2.56 | 176 |

*Table 3*: Summary of the optical transmittance characterization at room temperature: apparent optical band gap energy ($E_g^{Abs}$), absorption band edge broadening ($\Delta E$) and linear absorption well above the band gap ($\alpha_0$) of the samples under study.

The evolution of the band gap energy as function of the Al mole fraction of the $Al_xIn_{1-x}N$ layers is shown in Figure 7 (black squares). This evolution is strongly influenced by the doping levels of the studied films. For example, samples deposited by RF sputtering show higher carrier concentration ($\sim 10^{20}$ cm$^{-3}$) and band gap energy values for InN (1.76eV) and high-In content $Al_xIn_{1-x}N$ compared to samples grown by MBE (InN: $E_g$ = 0.7 eV), which present much lower residual carrier concentration ($\sim 10^{18}$ cm$^{-3}$). This difference in band gap energy is explained by the Burstein-Moss effect, which is reduced at higher Al contents due to the reduction of the carrier concentration (see Fig. 5). For comparison, Fig. 7 includes the results obtained by Nuñez-Cascajero *et al.* in layers deposited by RF sputtering under similar growth conditions (red line) [24], and the ones reported by Terashima *et al.* from samples grown by MBE (green line) [42].

From this relation, we can extract the bowing parameter (*b*), which is a non-linear parameter used in the modified Vegard´s law, $E_{gAl_xIn_{1-x}N} = xE_{gAlN} + (1-x)E_{gInN} - bx(1-x)$, that takes into account the deviation from linearity of the dependence of the band gap with the Al content. From the fitting shown in Fig. 7, we estimated a bowing parameter of *b* = 4.78 eV for our sample (black line), which is in good agreement with reported values for $Al_xIn_{1-x}N$ samples with similar doping levels [24].

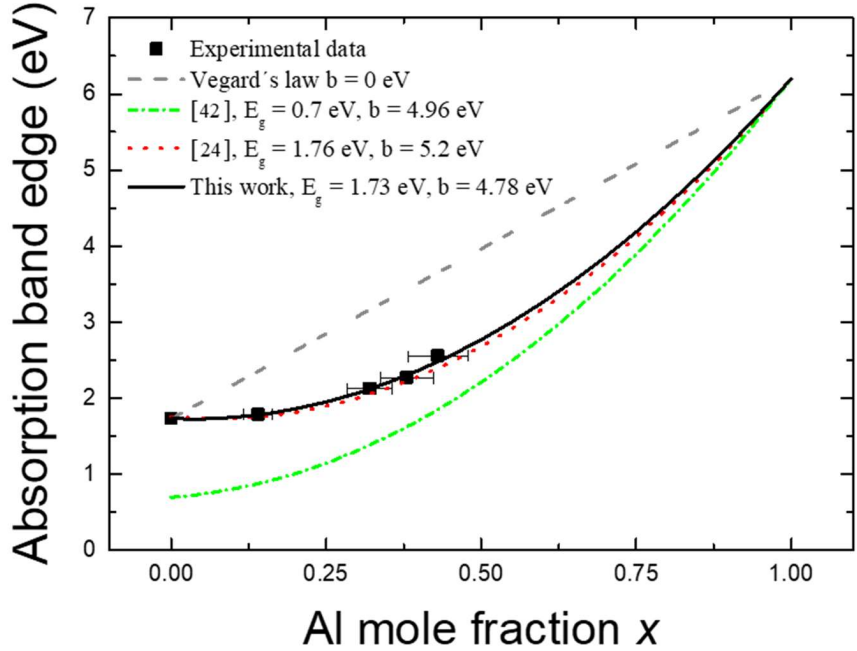

*Figure 7*: Band gap energy of the $Al_xIn_{1-x}N$ layers as a function of the Al content (black squares) and their fitting using modified Vergard's law (black line) to extract the bowing parameter. For comparison, results obtained by Nuñez-Cascajero *et al.* using the sputtering technique (red line) [24], and results reported by Terashima *et al.* for $Al_xIn_{1-x}N$ samples grown by molecular beam epitaxy with a pure InN band gap energy (green line) [42] are also presented. The grey line corresponds to Vegard's law assuming $b = 0$.

### 3.4.2 Photoluminescence measurements

Figure 8 shows the low-temperature (11K) PL emission of $Al_xIn_{1-x}N$ samples grown on (a) Si(100) and (b) sapphire. The dominant emission energy blue shifts from $E_{PL} = 1.6$ eV to $E_{PL} = 1.8$ eV when increasing the Al content up to $x = 0.36$ (see Table 4). No PL emission was observed for $Al_xIn_{1-x}N$ layers with higher Al content, whatever of the substrate. The FWHM of the PL emission varies from 390 to 510 meV depending on the sample, being in the same order than values obtained in similar sputtered samples [24].

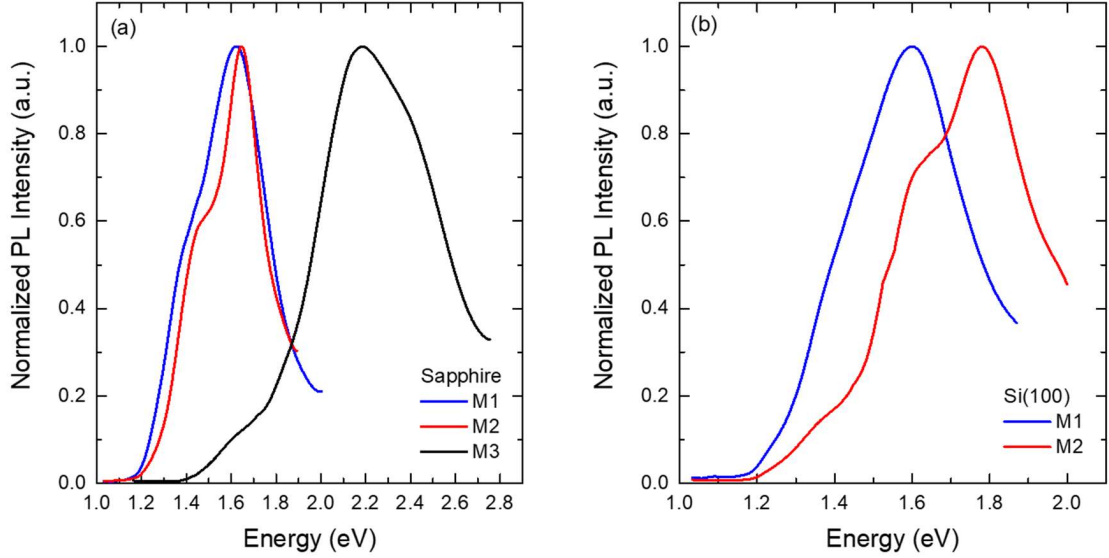

*Figure 8*: Normalized low-temperature (11 K) PL emission of $Al_xIn_{1-x}N$ layers deposited on: (a) sapphire substrates and (b) Si(100). For $x > 0.36$ no PL emission was observed on any substrate.

Table 4 shows the PL emission energy of the samples at room temperature (300 K). The value of the PL emission obtained for the M2 sample on Si(100) substrate at RT is close to value obtained by H. F. Liu, et. al. [36] for $Al_{0.28}In_{0.72}N$ on Si(100) samples.

| Sample | Substrate | Al content $x$ | $E_{PL}$ at 11 K (eV) | $E_{PL}$ at 300 K (eV) | Stokes shift (meV) |
|---|---|---|---|---|---|
| M1 | Si(100) | 0 | 1.59 | 1.56 | - |
| M2 | | 0.12 | 1.74 | 1.71 | - |
| M1 | Sapphire | 0 | 1.62 | 1.59 | 140 |
| M2 | | 0.14 | 1.62 | 1.61 | 180 |
| M3 | | 0.32 | 1.77 | 1.75 | 380 |

*Table 4*: Summary of the optical photoluminescence characterization: PL emission energy at 11K and 300 K and the Stokes shift for $Al_xIn_{1-x}N$ on Si(100) and sapphire samples.

From the obtained results for the samples grown on sapphire, we can extract the Stokes shift as the difference between the band gap energy, obtained from transmission measurements ($E_g^{Ab}$), and the PL emission energy ($E_{PL}$) at 300 K. The increase of the Stokes shift from 140 meV for InN to 380 meV for the $Al_{0.32}In_{0.68}N$ can be assigned to the alloy inhomogeneity or to defects/impurities present in the layer acting as undesired recombination centers. This high Stokes shift agrees with the calculations of Jiang *et. al.* [43].

All things considered, the XRD patterns obtained in this work for $Al_xIn_{1-x}N$ samples deposited by RF sputtering are similar across the literature, where the preferred growth orientation is along the *c*-axis and no phase separation is detected. The characteristic morphology with the presence of small grains in the surface for $x > 0$ is like the one presented in [13] for $Al_{0.17}In_{0.83}N$ on Si(111) samples; and differ from the ones reported in [21] for $Al_{0.3}In_{0.7}N$ on Si(100) samples, which are more columnar-like. Our layers own a carrier concentration in the same range of the ones reported in [13], and band gap energies in good agreement with the ones reported by [14] and [21] on $Al_xIn_{1-x}N$ ($x$ from 0 to 0.3) layers.

## 4. CONCLUSIONS

$Al_xIn_{1-x}N$ films with low-to-mid Al content ($x \sim 0$-$0.50$) were deposited via RF sputtering on different substrates, i.e., Si(100), Si(111) and sapphire, allowing the comparison of the obtained samples. The increase of the Al mole fraction improves the structural and morphological quality of the layers, achieving a minimum FWHM of the (0002) $Al_xIn_{1-x}N$ rocking curve of ~2.8º and a minimum rms surface roughness of ~2.4 nm for samples grown on both Si substrates with $x \sim 0.49$. Moreover, the $Al_xIn_{1-x}N$ layer morphology evolves from closely-packed columnar towards compact when increasing the Al content. The columnar nature of the films was further confirmed by TEM measurements, which showed comparable size and shape of the columnar grains for different silicon substrate orientations, in agreement with a similar defect density in the different samples (i.e., on silicon (111) and silicon (100)). On the other hand, $Al_xIn_{1-x}N$ layers deposited on sapphire present higher crystal quality, more compact morphology and lower rms surface roughness than the ones grown on Si regardless their Al content.

Hall-Effect measurements reveal a carrier concentration above $10^{20}$ cm$^{-3}$ for the Al$_x$In$_{1-x}$N layers with $x < 0.36$, probably induced by unintentional doping of the material during deposition. The optical band gap energy of the Al$_x$In$_{1-x}$N samples ranges from 1.75 eV (InN) to 2.56 eV ($x \sim 0.43$) depending on their Al content. Additionally, Al$_x$In$_{1-x}$N samples deposited on both Si and sapphire substrates exhibit a blue-shift of the PL emission energy when increasing the Al content from 1.59 eV (InN) to 1.75 eV ($x \sim 0.32$). In this work, we demonstrate the ability to produce high-quality Al$_x$In$_{1-x}$N layers on Si with low-to-mid Al content via RF sputtering regardless the substrate orientation.